\documentclass[runningheads]{llncs}

\usepackage[T1]{fontenc}
\usepackage{amsmath}
\usepackage[table,xcdraw]{xcolor}
\usepackage{multirow}
\usepackage[squaren,Gray]{SIunits}
\usepackage{hhline}
\usepackage{amsfonts}
\usepackage{tikz}
\usepackage{hyperref}
\usepackage{adjustbox}
\def\doi#1{\href{https://doi.org/\detokenize{#1}}{\url{https://doi.org/\detokenize{#1}}}}

\usepackage{multicol}
\usepackage{graphicx}
\usepackage{listings}
\begin{document}
\title{TriadNet: Sampling-free predictive intervals for lesional volume in 3D brain MR images}
\titlerunning{TriadNet: Predictive Intervals Estimation in 3D brain MRI}

\author{Benjamin Lambert \inst{1, 2} \and
Florence Forbes \inst{3} \and Senan Doyle \inst{2} \and Michel Dojat \inst{1}}
\authorrunning{B. Lambert et al.}

\institute{Univ. Grenoble Alpes, Inserm, U1216, Grenoble Institut Neurosciences, 38000, FR \and
Pixyl, Research and Development Laboratory, 38000 Grenoble, FR \and
Univ. Grenoble Alpes, Inria, CNRS, Grenoble INP, LJK, 38000 Grenoble, FR}
\maketitle              
\begin{abstract}
The volume of a brain lesion (e.g. infarct or tumor) is a powerful indicator of patient prognosis and can be used to guide the therapeutic strategy. Lesional volume estimation is usually performed by segmentation with deep convolutional neural networks (CNN), currently the state-of-the-art approach. However, to date, few work has been done to equip volume segmentation tools with adequate quantitative predictive intervals, which can hinder their usefulness  and acceptation in clinical practice. In this work, we propose \textbf{TriadNet}, a segmentation approach relying on a multi-head CNN architecture, which provides  both the lesion volumes and the associated predictive intervals simultaneously, in less than a second. We demonstrate its superiority over other solutions on BraTS 2021, a large-scale MRI glioblastoma image database.

\end{abstract}

\section{Introduction}
The lesional volume is a powerful and commonly used biomarker in brain MRI analysis and interpretation. Such an imaging biomarker is a guide  to predict the patient's neurological outcome in Stroke \cite{ghoneem2022association} or to assess the grade of a Glioblastoma \cite{baris2016role}. For Multiple Sclerosis (MS), the evolution of the lesional load between two patient's visits helps to assess the progress of the disease and to personalize his/her treatment \cite{mattiesing2022spatio} and even to predict the disability \cite{roca2020artificial}. For neurodegenerative diseases such as Alzheimer's disease, the brain atrophy is quantified by estimating the volume of different anatomical regions (e.g. hippocampus or amygdala) compared to normative values \cite{contador2021longitudinal}. 

Volume estimation is usually carried out through image segmentation, relying on Deep Convolutional Neural Networks (CNNs) trained on an annotated database, comprising both images and their corresponding manual delineations  \cite{hesamian2019deep}. CNNs provide a mask, which is generally correct for easy detectable regions or lesions, but whose accuracy may be more uncertain when the zone to segment is disputable even for an expert. 
To help clinicians to focus on the more subtle regions, we propose to associate quantitative Predictive Intervals (PIs) to volume estimation. Such PIs can straightforwardly be interpretated as uncertainty markers and facilitate the acceptance of advanced computerized tools by practitioners.

PI construction has been mainly studied in the context of 1D regression tasks \cite{kivaranovic2020adaptive,pearce2018high,tagasovska2019single} and applications in the context of medical image processing are very scarce. To compute PIs for lesion counting in 2D medical images, reference work proposes either a sampling approach or a regression model \cite{eaton2019easy}. In the former, several plausible and diverse segmentation masks are generated for the same input image, forming a distribution over the quantity of interest (e.g lesion volume or number), from which the mean and the standard deviation can be extracted to define a PI. This Uncertainty Quantification (UQ) methodology offers several variants to generate the  diverse set of predictions. Popular UQ methods regroup the Monte Carlo Dropout (MC) \cite{gal2016dropout}, Deep Ensemble \cite{lakshminarayanan2017simple}, or Test Time Augmentation (TTA) \cite{wang2019aleatoric}. Based on sampling, UQ methods are associated with an important computational burden to obtain the predictions. With the regression approach, a network is trained to directly predict the PI's components: the mean value as well as the lower and upper bounds from the data themselves. As no assumptions are made regarding the distribution of the regressed variable, this approach is referred to as Distribution-Free Uncertainty Quantification (DFUQ) \cite{pearce2018high}. In this direction, we introduce a sampling-free approach based on an original CNN architecture called TriadNet which exhibits the following assets: 
\begin{itemize}
    \item It enhances the 3D volume estimation with associated reliable PIs.
    \item It allows a fast and distribution-free estimation of PIs.
    \item The methodology is simple to implement and can be applied to any encoder-decoder segmentation architecture.
   
\end{itemize}

\section{Problem Definition}
We consider a 3D segmentation problem with $N$ classes. Excluding the background class, we aim at estimating the true unknown volumes $Y\in \mathbb{R}^{N-1}$ of each foreground classes based on the predicted segmentation. In this context, for an estimation $X$ of the volume, seen as a random variable, we define a predictive interval $\Gamma_{\alpha}(X)$ as a range of values that are conditionned to contain $Y$, the actual volume, with a certain degree of confidence $1-\alpha$ (e.g $90\%$ or $95\%$). That is,
given a series of estimated volumes $X_{1} \ldots X_n$ and their associated ground truth volumes $Y_1 \ldots Y_n$, $\Gamma_{\alpha}(\cdot)$ should be learned as to satisfy:
\begin{equation}
    P(Y_{\text{test}}\in\Gamma_{\alpha}(X_{\text{test}})) \geq 1-\alpha
\end{equation}
for any $(Y_{\text{test}},X_{\text{test}})$ following the same distribution as the $(Y_i,X_i)$'s.
This property is called the \emph{marginal coverage}, as the probability is marginal over the entire test dataset \cite{angelopoulos2021gentle}. 

Sampling-based PI estimation methods rely on the hypothesis that $X$ follows a normal distribution for each predicted class. Under this assumption, the mean value $\mu_X$ and standard deviation $\sigma_X$ of the distribution are estimated by sampling several distinct predictions for the same input, and PI are constructed as $\Gamma_{\alpha}(X) = [\mu_X - z\sigma_X, \mu_X + z\sigma_X]$
where $z$ is the number of standard deviation, stipulating the degree of confidence of the interval. For instance, for a $90\%$ confidence interval, $z$ corresponds to $1.65$. In contrast, direct PI estimation techniques (including regression approaches and our proposed TriadNet) directly output the mean value $\mu$, the lower bound $l_b $ and the upper bound $u_b$ ($l_b\leq\mu\leq u_b$), without sampling.

\section{Our solution: TriadNet}
\begin{figure}[ht!]
\centering
\includegraphics[width=\textwidth]{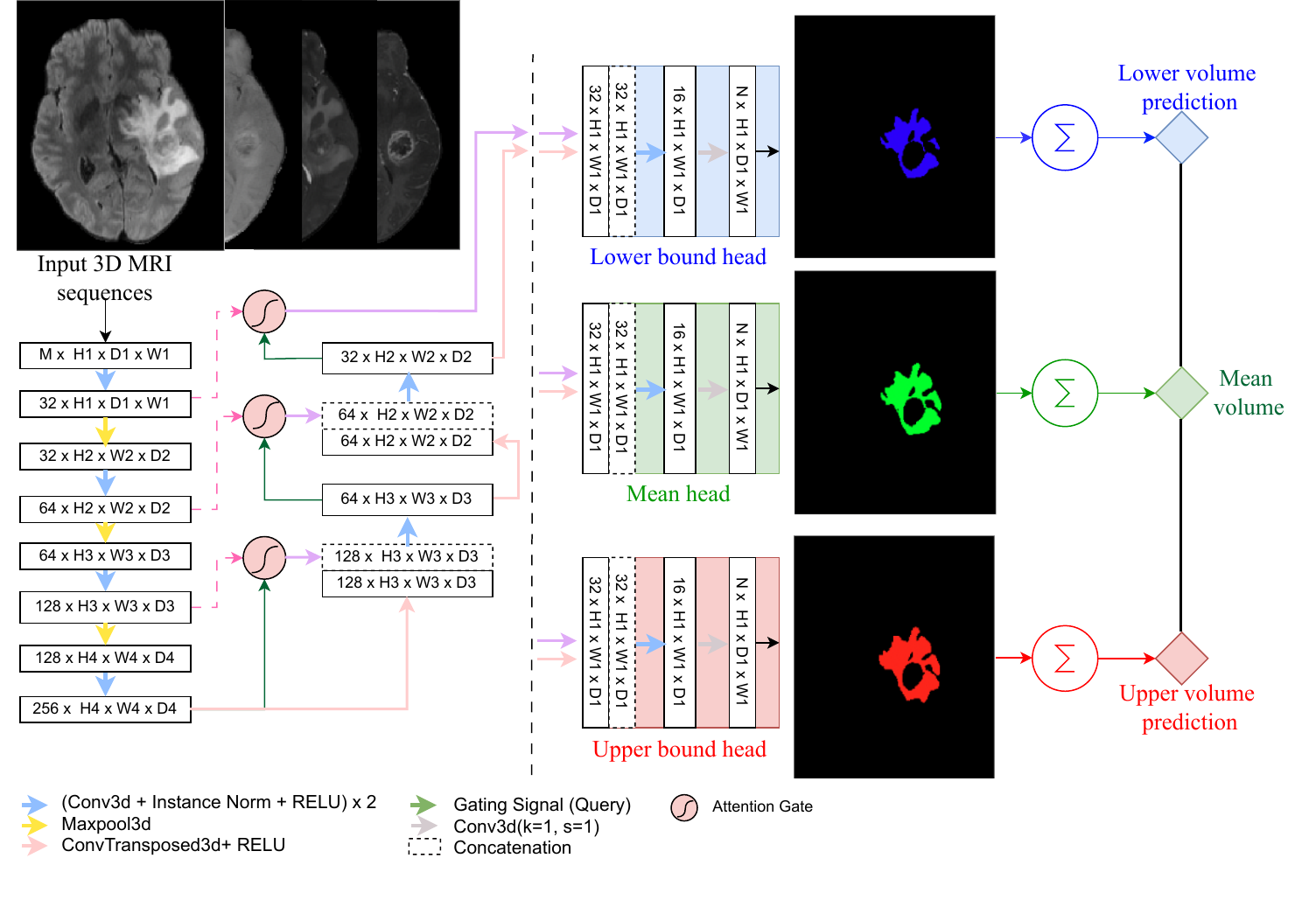}
\caption{The Triadnet architecture. Each head yields a distinct mask for each class:  lower bound, mean and upper bound masks. For ease of visualization, we only represent for a Glioblastoma application, the masks for the \emph{edematous} class.} \label{fig1}
\end{figure}

\emph{Overview:} TriadNet corresponds to a CNN model modified in order to produce three outputs for each segmented class: the mean, lower bound and upper bound masks (see Figure \ref{fig1}).  To obtain these distinct masks, we propose a multi-head architecture as well as a novel learning objective, the TriadLoss. The masks are then used to directly estimate the class-wise mean volume as well as the lower and upper bounds, by summing the segmented voxels.

\subsubsection{TriadNet: the architecture}\label{sec:Triadnet} relies on the Attention Unet 3d (AttUNet3d) \cite{oktay2018attention} as backbone. We modified it by duplicating the output convolutional block in order to obtain a total of 3 separate and identical heads. Each head generates a specific mask: respectively, one corresponding to the lower bound, one to the upper bound, and one for the mean value, by predicting a probabilistic distribution $p_{n,i}$ over the $N$ classes and for each voxel $i$. This modification only slighly increase the complexity of the segmentation model, raising the number of parameters from 5 millions to 5.3 millions.

\subsubsection{TriadLoss: the objective function}\label{sec:Triadloss}
is built on the observation that the lower bound mask should be more restrictive (\emph{i.e.} higher precision and lower recall) than the mean mask. Similarly, the upper bound mask should be more permissive (\emph{i.e.} higher recall and lower precision). To achieve this, we propose to rely on the Tversky loss \cite{salehi2017tversky}, which provides a direct control on the trade-off between recall and precision. The Tversky loss $T_{\alpha, \beta}$ is an extension of the popular Dice loss \cite{milletari2016v}, with 2 extra hyperparameters $\alpha$ and $\beta$ which respectively control the weighting of False Positives (FP) and False Negatives (FN). With $\alpha=\beta=0.5$, the Tversky loss is strictly equivalent to the standard Dice loss. 

Writing $p_{\textit{lower}}$, $p_{\textit{mean}}$ and $p_{\textit{upper}}$ the outputs of each head and $y$ the ground-truth segmentation, we defined the Triad loss as: 

\begin{equation}
    \text{TriadLoss} = T_{1-\gamma, \gamma}(p_{\textit{lower}, y}) + T_{0.5, 0.5}(p_{\textit{mean}, y}) + T_{\gamma, 1-\gamma}(p_{\textit{upper}, y})
\end{equation}

with $\gamma$ an hyperparameter in the range $]0, 0.5[$ controlling the penalties applied to FP and FN during the training of the lower and upper bound heads. In other words, the mean decoder was trained with a standard Dice Loss. To obtain more restrictive masks (and lower volumes), the lower bound decoder was trained to minimize FP at the expense of a higher FN rate. Similarly, to obtain more permissive masks (and larger volumes), the upper bound decoder sought to minimize FN at the expense of a higher number of FP. 

\section{Material and Methods}
\subsection{Datasets}
We illustrate our framework on a brain tumor segmentation task, using the open-source part of the BraTS 2021 dataset \cite{baid2021rsna} containing 1251 patients. Four MRI sequences are available for each patient: FLAIR, T1, T2, and  T1ce (T1w with contrast agent). The ground truth segmentation masks contain 4 classes: the background, the necrotic tumor core, the edematous, and the GD-enhancing (GDE) tumor. We randomly split the data into a training fold (651), a calibration fold (200), and a testing fold (400).

\subsection{Comparison with known approaches}
We compared TriadNet with 3 sampling-based approaches: Confidence Thresholding (CT), Monte Carlo dropout (MC), and Test Time Augmentation 
(TTA), as well as a sampling-free PI estimation framework based on the training of a regression CNN (RegCNN). 

\textbf{Confidence Thresholding (CT)} is a simple approach to obtain PI's from the output probability estimates produced by a trained segmentation model. For each class, the probability map is binarized with progressively increasing thresholds. As the threshold increases, fewer voxels are segmented, thus the volume decreases. As this method relies on the calibration of the output probabilities, we perform Temperature Scaling \cite{guo2017calibration} on the trained segmentation model before performing CT. 

\textbf{Monte Carlo Dropout (MC)} is based on the Dropout technique \cite{srivastava2014dropout} which consists of turning a subset of the model parameters off, to prevent overfitting. The MC dropout technique proposes to keep dropout activated during inference, meaning that $T$ forward steps of the same image through the MC dropout model will lead to $T$ different segmentations (and thus volume estimates), as the dropout mask is randomly sampled at each step. 

\textbf{Test Time Augmentation (TTA)} consists in using data
augmentation to generate alternative versions of the input images. Each augmented image is processed by the segmentation model, yielding to a distinct estimation of the volumes. By repeating this process, a distribution over volumes can be obtained, from which the PI is derived. 

\textbf{Regression CNN} (RegCNN) proposes to train a regression neural network to directly predict the lower, mean and upper bounds of the target quantity from the data itself \cite{angelopoulos2021gentle,eaton2019easy}. To achieve this, the Pinball loss $P_t$ can be used to train the model to predict a desired quantile $t$. In our study, the regressor took as input the MRI sequences and automated segmentation produced by a segmentation model and was trained to predict three scores for each segmentation class, namely the $q_{\alpha/2}$, $q_{0.5}$ and $q_{1-\alpha/2}$ quantiles, allowing the construction of $(1-\alpha)\%$ confidence intervals. To do so, the regressor was trained with a compound loss $L=P_{\alpha/2} + P_{0.5} + P_{1-\alpha/2}$ to learn each quantile. 

\subsection{Post-hoc PI calibration}
In practice, the predicted PIs may be inaccurate and not respect the desired \emph{marginal coverage} property. To alleviate this, PI post-hoc calibration is usually performed using a set-aside calibration dataset \cite{angelopoulos2021gentle}. This calibration step aims at finding the optimal corrective value $q$ such that the calibrated PIs achieve the desired $(1-\alpha)\%$ coverage on the calibration dataset. 

In the case of sampling-based PI, the corrective value takes the form of a multiplicative factor applied to the standard deviation (Equation \ref{eq:sampling}). Alternatively, if the PI estimation is direct, $q$ corresponds to an additive factor applied to the lower and upper bounds (Equation \ref{eq:direct}) :
\begin{align}
    \Gamma_{\alpha,\text{cal}}(X)&=[\mu_X - q\sigma_X, \mu_X + q\sigma_X] \label{eq:sampling}  \\
    \Gamma_{\alpha,\text{cal}}(X)&=[l_b - q, u_b + q]  \label{eq:direct} 
\end{align}

\subsection{Evaluation}
We performed all our experiments with $\alpha=0.1$, meaning that we focussed on $90\%$ PIs. Segmentation performance was assessed using the Dice score (DSC) between the predicted segmentation and the ground truth delineations (for TriadNet, the Dice was computed using the \emph{mean} predicted mask). We also used the Mean Average Error (MAE) between the estimated mean volumes and the true volumes to assess the reliability of the volume prediction. 

Useful PIs should have two properties. They should i) achieve the desired \emph{marginal coverage} and ii) be the narrowest possible in order to be informative. To verify this, we computed two scores for PIs: the coverage error ($\Delta f$) and the interval width ($W$). $\Delta f$ is defined as the distance between the empirical coverage and target coverage ($90\%$). $W$ is the average distance between the lower and upper bounds. Note that a successful PI calibration should ensure $\Delta f \geq 0$. However, as the width of intervals tend to augment with $\Delta f$, a value close to 0 is preferred. To estimate computational efficiency,  we also reported the average time to produce a segmentation and PI for one input MRI volume.

To assess the impact of the choice of the $\gamma$ hyper-parameter in the TriadLoss on PI quality, we trained Triadnet models with varying $\gamma$ values, ranging from $0.1$ to $0.4$. To obtain robust statistics, each model is trained 5 times and we reported the average and standard deviation for each metrics.

\subsection{Implementation Details}
Three types of segmentation models are used in this study. First, \emph{Baseline} AttUnet3d was trained to serve as a common basis for the implementation of CT, TTA and RegCNN approaches. For MC, we trained a dedicated \emph{Dropout} AttUnet3ds by adding a dropout rate of $20\%$ in each layer of the encoder and decoder. The last type of segmentation model was our proposed TriadNet. All models were trained with the ADAM optimizer \cite{kingma2014adam}, with a learning rate of $2e-4$, using the Dice loss for \emph{Baseline} and \emph{Dropout} models and the TriadLoss for TriadNet. For CT-based PIs, we used $20$ different thresholds uniformly distributed in the range $[0.01, 0.99]$ to binarize the probability maps. For MC dropout, we performed $T=20$ forward passes of the same input image with dropout activated to obtain the PIs. To implement the TTA baseline, we generated $20$ random augmentations for each input MRI using flipping, rotation, translation and contrast augmentation with randomized parameters, implemented using the TorchIO Data Augmentation library \cite{perez2021torchio}. Finally for RegCNN, we used an open-source regressor CNN implementation \footnote{\url{https://docs.monai.io/en/stable/_modules/monai/networks/nets/regressor.html}} \cite{cardoso2022monai}.

\section{Results and Discussion}
\begin{figure}
    \centering
    \includegraphics[width=\textwidth]{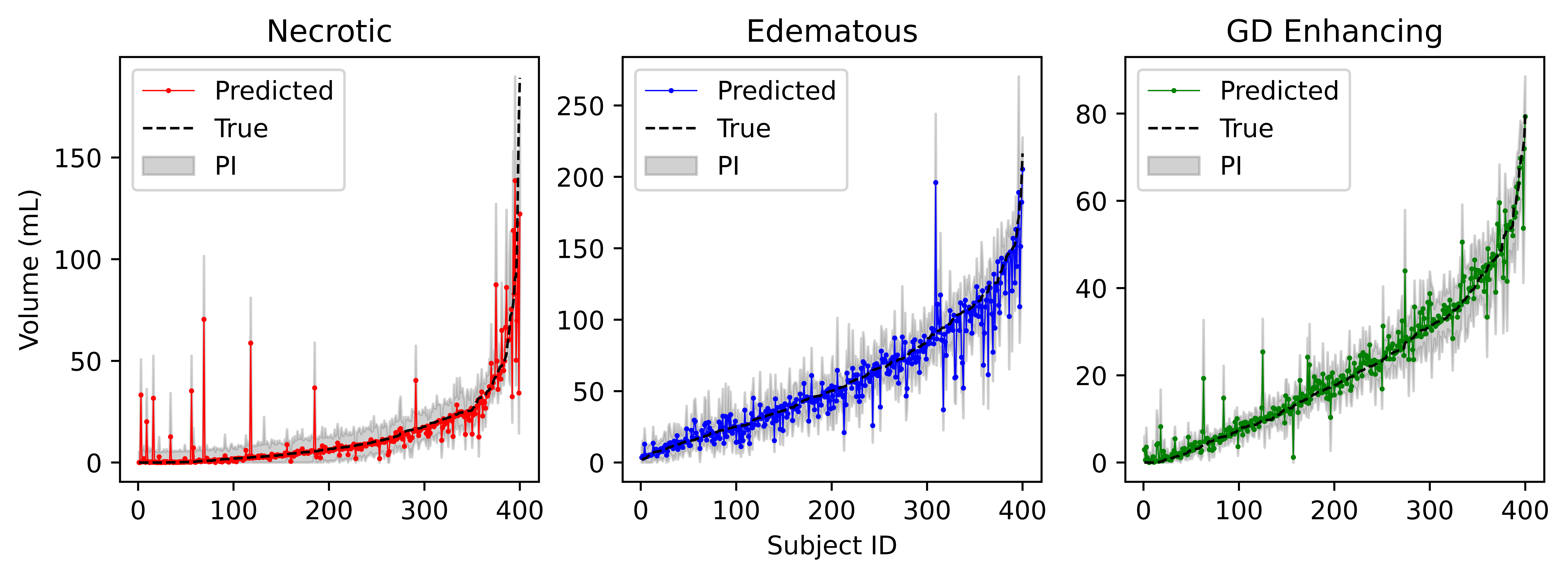}
    \caption{Predictive intervals generated by TriadNet ($\gamma=0.2$) on the test dataset.}
    \label{fig:pi_illustration}
\end{figure}

Table \ref{tab1} presents the performance of segmentation (DSC) and PIs for each approach and for all 3 segmented tumor tissues; and Table \ref{table2}, the average computation time for each method. Finally, Figure \ref{fig:pi_illustration} provides an illustration of PI computed by our proposed TriadNet on the test dataset. 

\setlength{\tabcolsep}{4.5pt}
\begin{table}[ht!]
\caption{Performances for each tumor tissue for each method. $\Delta f$: coverage error, $W$: average interval width. Mean scores obtained over 5 runs. SD: standard deviation.  }\label{tab1}
\centering
\begin{tabular}{clcccc}
\hline
& Method & \begin{tabular}[c]{@{}c@{}} $\Delta f$ \\ ($\%\pm$SD) \end{tabular} & \begin{tabular}[c]{@{}c@{}} W $\downarrow$ \\ ($mL\pm$SD) \end{tabular} & \begin{tabular}[c]{@{}c@{}} MAE $\downarrow$ \\ ($mL\pm$SD) \end{tabular} & \begin{tabular}[c]{@{}c@{}} DSC $\uparrow$ \\ ($\pm$SD) \end{tabular} \\ \hline

\parbox[t]{2mm}{\multirow{8}{*}{\rotatebox[origin=c]{90}{Necrotic}}} & CT & 5.6 $\pm$ 1.5 & 32.3 $\pm$ 5.3 & 3.4 $\pm$ 0.1 & \textbf{0.76 $\pm$ 0.00} \\
& TTA   & 6.3 $\pm$ 1.1 & 25.0 $\pm$ 3.7 & 3.5 $\pm$ 0.1 & \textbf{0.76 $\pm$ 0.00} \\
& RegCNN  & 6.1 $\pm$ 1.7 & 25.8 $\pm$ 5.7 & 6.3 $\pm$ 3.6 &  \textbf{0.76 $\pm$ 0.00} \\
& MC dropout    & 5.6 $\pm$ 0.6 & 20.6 $\pm$ 2.1 & 3.3 $\pm$ 0.1 & \textbf{0.76 $\pm$ 0.00} \\
& TriadNet ($\gamma=0.1)$ & 4.5 $\pm$ 0.8 & 14.5 $\pm$ 1.4 & 3.4 $\pm$ 0.1 & 0.75 $\pm$ 0.00 \\ 
& TriadNet ($\gamma=0.2)$ & \textbf{3.4 $\pm$ 0.6} &\textbf{ 13.7 $\pm$ 1.1 } & \textbf{3.2 $\pm$ 0.1} & \textbf{0.76 $\pm$ 0.00} \\ 
& TriadNet ($\gamma=0.3)$ & 4.1 $\pm$ 0.9 & 15.0 $\pm$ 0.4 & 3.3 $\pm$ 0.1 & \textbf{0.76 $\pm$ 0.01} \\ 
& TriadNet ($\gamma=0.4)$ & 4.4 $\pm$ 0.4 & 16.6 $\pm$ 0.8 & 3.4 $\pm$ 0.1 & 0.75 $\pm$ 0.00 \\ \hline

\parbox[t]{2mm}{\multirow{8}{*}{\rotatebox[origin=c]{90}{Edematous}}} & CT    & 1.4 $\pm$ 1.2 & 54.4 $\pm$ 12.3 & 8.2 $\pm$ 0.9 & \textbf{0.85 $\pm$ 0.01} \\
& TTA   & -1.3 $\pm$ 2.3 & 34.9 $\pm$ 2.5 &  7.7 $\pm$ 0.2 & \textbf{0.85 $\pm$ 0.01} \\
& RegCNN  & 1.7 $\pm$ 0.9 & 41.9 $\pm$ 1.7 & 9.2 $\pm$ 0.4 & \textbf{0.85 $\pm$ 0.01} \\
& MC dropout   & \textbf{-0.01 $\pm$ 1.4} & 32.0 $\pm$ 2.0 & 7.5 $\pm$ 0.2 & 0.84 $\pm$ 0.01 \\
& TriadNet ($\gamma=0.1)$ & 0.9 $\pm$ 0.7 & 31.8 $\pm$ 0.8 & 7.4 $\pm$ 0.5 & \textbf{0.85 $\pm$ 0.00} \\ 
& TriadNet ($\gamma=0.2)$ & 1.6 $\pm$ 1.2 & \textbf{30.2 $\pm$ 1.7} & 7.2 $\pm$ 0.2 & \textbf{0.85 $\pm$ 0.00} \\ 
& TriadNet ($\gamma=0.3)$ & 3.2 $\pm$ 1.6 & 35.7 $\pm$ 4.3 & \textbf{7.1} $\pm$ 0.2 & \textbf{0.85 $\pm$ 0.00} \\ 
& TriadNet ($\gamma=0.4)$ & 1.4 $\pm$ 0.7 & 31.1 $\pm$ 2.2 & 7.5 $\pm$ 0.3 & 0.84 $\pm$ 0.01 \\ \hline
\parbox[t]{2mm}{\multirow{8}{*}{\rotatebox[origin=c]{90}{GDE}}}
& CT    & 3.6 $\pm$ 1.4 & 17.8 $\pm$ 3.5 & 2.0 $\pm$ 0.0 & \textbf{0.85 $\pm$ 0.01} \\
& TTA   & 3.0 $\pm$ 1.8 & 10.6 $\pm$ 1.0  & 2.0 $\pm$ 0.1 & \textbf{0.85 $\pm$ 0.01} \\
& RegCNN  & \textbf{0.7 $\pm$ 0.5} & 22.2 $\pm$ 5.9 & 3.2 $\pm$ 0.3 & \textbf{0.85 $\pm$ 0.01} \\
& MC dropout   & 3.5 $\pm$ 1.3 & 10.0 $\pm$ 0.5  & 1.9 $\pm$ 0.1 & \textbf{0.85 $\pm$ 0.01} \\
& TriadNet ($\gamma=0.1)$ & 3.7 $\pm$ 1.1 & 11.0 $\pm$ 0.6 & \textbf{1.8 $\pm$ 0.1} & \textbf{0.85 $\pm$ 0.00} \\ 
& TriadNet ($\gamma=0.2)$ & 4.2 $\pm$ 1.4 & \textbf{8.9 $\pm$ 0.5} & \textbf{1.8 $\pm$ 0.1} & \textbf{0.85 $\pm$ 0.00} \\ 
& TriadNet ($\gamma=0.3)$ & 4.1 $\pm$ 0.4 & 9.3 $\pm$ 0.6 & \textbf{1.8 $\pm$ 0.1}& \textbf{0.85 $\pm$ 0.00} \\ 
& TriadNet ($\gamma=0.4)$ & 4.0 $\pm$ 0.5 & 11.0 $\pm$ 0.4 & 1.9 $\pm$ 0.1 & \textbf{0.85 $\pm$ 0.00} \\ 
\end{tabular}
\end{table}

\setlength{\tabcolsep}{4pt}
\begin{table}[t!]
\caption{Average prediction time to obtain a segmentation of a 3D MRI volume associated to predictive intervals on the volumes. SD=standard deviation}\label{table2}
\footnotesize
\centering
\begin{tabular}{lccccc}
         & CT  & TTA & RegCNN & MC & TriadNet  \\ \hline
Time (s$\pm$SD) $\downarrow$  & 1.1 $\pm$ 0.1 & 14.3 $\pm$ 1.2 & \textbf{0.3 $\pm$ 0.1} & 6.4 $\pm$ 0.1 & 0.6 $\pm$ 0.1  \\ 
\end{tabular}
\end{table}

Most methods provide PIs that, after calibration, achieve the target \emph{marginal coverage} property ($\Delta f \geq 0$). In terms of interval width (W), the narrowest intervals are provided by our proposed TriadNet parameterized by $\gamma=0.2$, while MC dropout ranks as the second best approach. To estimate the significance of this result, two-sided paired t-test between both methods were performed, showing that TriadNet's PI are significantly narrower compared to MC dropout ones ($p < 0.05$ for each tumor class). The best volume estimation, computed using the MAE, is also obtained by TriadNet, while RegCNN estimation is systematically the worst. In terms of segmentation quality (DSC scores), all models achieve very similar performances. Finally, regarding computational efficiency (Table 2),  RegCNN appears as the fastest approach, followed by TriadNet, both producing segmentation and associated PIs in less than one second for an input MRI volume. As expected, sampling approaches are much more time-consuming, with MC and TTA being respectively $10$ and $24$ times slower than our proposed TriadNet. 

The choice of the $\gamma$ parameter in the TriadLoss has proved to be important, with an optimal PI quality reached for $\gamma=0.2$, equivalent to a weighting of $0.8$ for FP and $0.2$ for FN in the lower bound head; and $0.2$ for FP and $0.8$ for FN in the upper bound head. This setting allows the different masks (lower, mean and upper) to be different enough to allow a reliable PI estimation, which is not the case with higher $\gamma$ values ($\gamma=0.3$ and $\gamma=0.4$). However when $\gamma$ is lower ($\gamma=0.1$), the penalty on FP and FN is too small, which yields to a larger amount of erroneous predictions, lowering PI quality. 

\section{Conclusion}
In this work, we addressed the problem of constructing PIs associated to 3D brain MR segmented volumes. Our proposed TriadNet provides narrower and thus more informative intervals in practice compared to competing methods, while preserving the desired \emph{marginal coverage} property. Interestingly, it is also 10 times faster than the second best baseline, MC dropout, making it suitable for clinical routine. Finally, it only requires a minor modification of the segmentation architecture, which has no negative impact on segmentation quality. Future work will investigate the robustness of TriadNet's predictive intervals in the presence of domain shift, and evaluate how our approach behaves with respect to the size of the target region, ranking from very small targets (e.g the hippocampus region or MS lesions) to very large (e.g the overall grey matter volume).

%
%

\bibliographystyle{splncs04}
\bibliography{biblio}

\begin{thebibliography}{10}
\providecommand{\url}[1]{\texttt{#1}}
\providecommand{\urlprefix}{URL }
\providecommand{\doi}[1]{https://doi.org/#1}

\bibitem{angelopoulos2021gentle}
Angelopoulos, A.N., Bates, S.: A gentle introduction to conformal prediction
  and distribution-free uncertainty quantification. arXiv preprint
  arXiv:2107.07511  (2021)

\bibitem{baid2021rsna}
Baid, U., Ghodasara, S., et~al.: The rsna-asnr-miccai brats 2021 benchmark on
  brain tumor segmentation and radiogenomic classification. arXiv preprint
  arXiv:2107.02314  (2021)

\bibitem{baris2016role}
Baris, M.M., Celik, A.O., et~al.: Role of mass effect, tumor volume and
  peritumoral edema volume in the differential diagnosis of primary brain tumor
  and metastasis. Clinical neurology and neurosurgery  \textbf{148},  67--71
  (2016)

\bibitem{cardoso2022monai}
Cardoso, M.J., Li, W., Brown, R., Ma, N., Kerfoot, E., Wang, Y., Murrey, B.,
  Myronenko, A., Zhao, C., Yang, D., et~al.: Monai: An open-source framework
  for deep learning in healthcare. arXiv preprint arXiv:2211.02701  (2022)

\bibitem{contador2021longitudinal}
Contador, J., P{\'e}rez-Mill{\'a}n, A., et~al.: Longitudinal brain atrophy and
  csf biomarkers in early-onset alzheimer’s disease. NeuroImage: Clinical
  \textbf{32},  102804 (2021)

\bibitem{eaton2019easy}
Eaton-Rosen, Z., Varsavsky, T., et~al.: As easy as 1, 2... 4? uncertainty in
  counting tasks for medical imaging. Medical Image Computing and Computer
  Assisted Intervention (MICCAI) 2019 pp. 356--364 (2019)

\bibitem{gal2016dropout}
Gal, Y., Ghahramani, Z.: Dropout as a {B}ayesian approximation: Representing
  model uncertainty in deep learning. International Conference on Machine
  Learning  \textbf{48},  1050--1059 (2016)

\bibitem{ghoneem2022association}
Ghoneem, A., Osborne, M.T., et~al.: Association of socioeconomic status and
  infarct volume with functional outcome in patients with ischemic stroke. JAMA
  Network Open  \textbf{5}(4),  e229178--e229178 (2022)

\bibitem{guo2017calibration}
Guo, C., Pleiss, G., Sun, Y., Weinberger, K.Q.: On calibration of modern neural
  networks. In: International conference on machine learning. pp. 1321--1330.
  PMLR (2017)

\bibitem{hesamian2019deep}
Hesamian, M.H., Jia, W., et~al.: Deep learning techniques for medical image
  segmentation: achievements and challenges. Journal of digital imaging
  \textbf{32},  582--596 (2019)

\bibitem{kingma2014adam}
Kingma, D.P., Ba, J.: Adam: {A} method for stochastic optimization. 3rd
  International Conference on Learning Representations, {ICLR} 2015  (2015)

\bibitem{kivaranovic2020adaptive}
Kivaranovic, D., Johnson, K.D., Leeb, H.: Adaptive, distribution-free
  prediction intervals for deep networks. International Conference on
  Artificial Intelligence and Statistics pp. 4346--4356 (2020)

\bibitem{lakshminarayanan2017simple}
Lakshminarayanan, B., Pritzel, A., Blundell, C.: Simple and scalable predictive
  uncertainty estimation using deep ensembles. Advances in Neural Information
  Processing Systems 30 pp. 6402--6413 (2017)

\bibitem{mattiesing2022spatio}
Mattiesing, R.M., Gentile, G., et~al.: The spatio-temporal relationship between
  white matter lesion volume changes and brain atrophy in clinically isolated
  syndrome and early multiple sclerosis. NeuroImage: Clinical  \textbf{36},
  103220 (2022)

\bibitem{milletari2016v}
Milletari, F., Navab, N., Ahmadi, S.A.: V-net: Fully convolutional neural
  networks for volumetric medical image segmentation. 2016 fourth international
  conference on 3D vision (3DV) pp. 565--571 (2016)

\bibitem{oktay2018attention}
Oktay, O., Schlemper, J., et~al.: Attention u-net: Learning where to look for
  the pancreas. Medical Imaging with Deep Learning (MIDL)  (2018)

\bibitem{pearce2018high}
Pearce, T., Brintrup, A., et~al.: High-quality prediction intervals for deep
  learning: A distribution-free, ensembled approach. International Conference
  on Machine Learning pp. 4075--4084 (2018)

\bibitem{perez2021torchio}
P{\'e}rez-Garc{\'\i}a, F., Sparks, R., Ourselin, S.: Torchio: a python library
  for efficient loading, preprocessing, augmentation and patch-based sampling
  of medical images in deep learning. Computer Methods and Programs in
  Biomedicine  \textbf{208},  106236 (2021)

\bibitem{roca2020artificial}
Roca, P., Attye, A., et~al.: Artificial intelligence to predict clinical
  disability in patients with multiple sclerosis using flair mri. Diagnostic
  and Interventional Imaging  \textbf{101}(12),  795--802 (2020)

\bibitem{salehi2017tversky}
Salehi, S.S.M., et~al.: Tversky loss function for image segmentation using 3d
  fully convolutional deep networks. Machine Learning in Medical Imaging (MLMI)
  Workshop, Held in Conjunction with MICCAI 2017 pp. 379--387 (2017)

\bibitem{srivastava2014dropout}
Srivastava, N., Hinton, G., et~al.: Dropout: a simple way to prevent neural
  networks from overfitting. The journal of machine learning research
  \textbf{15}(1),  1929--1958 (2014)

\bibitem{tagasovska2019single}
Tagasovska, N., Lopez-Paz, D.: Single-model uncertainties for deep learning.
  Advances in Neural Information Processing Systems  \textbf{32} (2019)

\bibitem{wang2019aleatoric}
Wang, G., Li, W., Aertsen, M., Deprest, J., Ourselin, S., Vercauteren, T.:
  Aleatoric uncertainty estimation with test-time augmentation for medical
  image segmentation with convolutional neural networks. Neurocomputing
  \textbf{338},  34--45 (2019)

\end{thebibliography}

\end{document}